\documentclass[letterpaper,twocolumn,10pt]{article}
\usepackage{usenix-2020-09}

\usepackage{url}

\usepackage{breakurl}
\usepackage{hyperref}

\usepackage{tikz}
\usepackage{amsmath}
\usepackage{multirow}
\usepackage{booktabs}
\usepackage{xspace}
\usepackage{titling}
\posttitle{\par\end{center}\vspace{-40pt}}

\RequirePackage{xcolor}
\definecolor{RED}{rgb}{1,0,0}
\definecolor{BLUE}{rgb}{0,0,1}
\definecolor{White}{rgb}{1,1,1}

\usepackage{paralist}

\newcommand{\mycaption}[2]{\caption{\textbf{#1}. {#2}}}
\newcommand{\sref}[1]{\S\ref{#1}}
\newcommand{\vheading}[1]{\vspace{0.05in}\noindent\textbf{#1}}
\newcommand{\viheading}[1]{\vspace{0.05in}\noindent\emph{#1}}
\newcommand{\fsync}{\texttt{fsync()}\xspace}

\newcommand{\eg}{\textit{e.g.,}\xspace}

\newcommand{\myx}{$\times$\xspace}
\newcommand{\vtt}[1]{\texttt{#1}\xspace}

\newcommand{\sysname}{{\textsc{FlyTrap}}\xspace}

 \newcommand{\sysfsync}{\texttt{fsync()}\xspace}

\newcounter{mylabelcounter}

\makeatletter
\newcommand{\labelText}[2]{%
#1\refstepcounter{mylabelcounter}%
\immediate\write\@auxout{%
  \string\newlabel{#2}{{1}{\thepage}{{\unexpanded{#1}}}{mylabelcounter.\number\value{mylabelcounter}}{}}%
}%
}
\makeatother

\newcommand{\uniquebugs}{18\xspace}
\newcommand{\totalbugs}{18\xspace}
\newcommand{\fixedbugs}{12\xspace}

\newcounter{bugcounter}
\newenvironment{bug}[1][\thebugcounter]{\refstepcounter{bugcounter}\par\noindent{\thebugcounter}}{}

\newcounter{observationcounter}
\newenvironment{observation}[1][]{\refstepcounter{observationcounter}
   \vspace{5pt} \noindent \textbf{Observation~\theobservationcounter: #1} \rmfamily}{}
   
\newcounter{lessoncounter}
\newenvironment{lesson}[1][]{\refstepcounter{lessoncounter}
   \vspace{5pt} \noindent \textbf{Lesson~\thelessoncounter: #1} \rmfamily}{}
   
\usepackage{tikz}
\usepackage{pgfplots}
\usetikzlibrary{calc,trees,positioning,arrows,chains,shapes.geometric,%
  decorations.pathreplacing,decorations.pathmorphing,decorations.text,shapes,%
  matrix,shapes.symbols,patterns,shadows}

\newcommand{\ccfuzzer}{Syzkaller\xspace}

\newcommand{\plot}[2]{
  \pgfplotstableread[col sep=comma]{#1}\datatable
  \begin{tikzpicture}
    \begin{axis}[
        ymode=log,
        title={#2},
        ymajorgrids,
        tick style={draw=none},
        y axis line style={draw opacity=0},
        log ticks with fixed point, 
        table/col sep=comma,
        tick label style={font=\footnotesize},
        xlabel=\# of bugs found,
        ylabel=Time Taken (s),
        xtick=data,
        extra y ticks = {1},
        extra y tick labels = {1},
        xticklabels from table={\datatable}{Bugs},
        legend entries={Syzkaller,ACE},
        legend style={draw=none},
        every axis legend/.append style={at={(0,1)}, anchor=north west, outer xsep=5pt, outer ysep=5pt,},
        ymin=1,
        ymax=1000000
      ]
      \addplot+ [
        thick,
      ]
      table[
        x expr=\coordindex,
        x = Bugs,
        y = Syzkaller,
        col sep=comma]{#1};
       
       \addplot+ [
          thick,
       ]
        table[
        x expr=\coordindex,
        x = Bugs,
        y = ACE,
        col sep=comma]{#1};
    \end{axis}
  \end{tikzpicture}
}

\hypersetup{linkcolor=blue}

\begin{document}

\date{}

\title{\Large \bf Finding and Analyzing Crash-Consistency Bugs\\ in Persistent-Memory File Systems\vspace{1cm}}

\author{
{\rm Hayley LeBlanc}\\
\emph{University of Texas at Austin}
\and
{\rm Shankara Pailoor}\\
\emph{University of Texas at Austin}
\and
{\rm Isil Dillig}\\
\emph{University of Texas at Austin}
\and
{\rm James Bornholt}\\
\emph{University of Texas at Austin}
\and
\rm{Vijay Chidambaram}\\
\emph{University of Texas at Austin and VMware Research}
} 

\maketitle

\begin{abstract}
  We present a study of crash-consistency bugs in persistent-memory (PM) file systems and analyze their implications for file-system design and testing crash consistency.
  We develop \sysname, a framework to test PM file systems for crash-consistency bugs.
  \sysname discovered \totalbugs new bugs across four PM file systems;
  the bugs have been confirmed by developers and many have been already fixed.
  The discovered bugs have serious consequences such as breaking the atomicity of rename or making the file system unmountable.
  We present a detailed study of the bugs we found and discuss some important lessons from these observations.
  For instance, one of our findings is that many of the bugs are due to logic errors, rather than errors in using flushes or fences;
  this has important applications for future work on testing PM file systems.
  Another key finding is that many bugs arise from attempts to improve efficiency by performing metadata updates in-place and that recovery code that deals with rebuilding in-DRAM state is a significant source of bugs.
  These observations have important implications for designing and testing PM file systems.
  Our code is available at \url{https://github.com/utsaslab/flytrap}.
\end{abstract} 

\section{Introduction}
\label{sec-intro}

Persistent memory (PM) is a new storage-class memory technology
that offers extremely low-latency persistent storage
and fine-grained access to storage over the memory bus~\cite{swanson-pm, pm-arxiv}.
PM technology has long been a focus of research,
and more recently has been commercialized by Intel~\cite{optane}.
A number of file systems~\cite{Xu2016, Condit2009, Dulloor2014, aerie, Kadekodi19, Kadekodi21, Kwon17, Dong2019, Anderson20} have been developed that exploit PM's advantages
to build faster, safer storage systems.

One of the main responsibilities of a file system is to keep the user's data safe in the event of a crash due to a power loss or a kernel bug.
To do so, the file system should be \emph{crash consistent}: it should recover after the failure to a consistent state without losing data the user expected to be persistent~\cite{GangerPatt-Metadata94, ChidambaramPhd15, crash-cacm15, bornholt:ferrite}.
However, there are no open-source tools that can test whether a PM file system is crash consistent without annotating or modifying the file system.
Existing crash consistency testing tools like CrashMonkey~\cite{Mohan18} or Hydra~\cite{Kim2019} are not compatible with the unique storage stack of PM file systems where the media is directly accessed using processor load and store instructions at fine granularity.

This paper makes two contributions.
First, it presents \sysname, a framework for testing the crash consistency of PM file systems (\sref{sec:tools}).
Given a workload, and a target PM file system, \sysname simulates crashes at different points in the workload, creating \emph{crash states} reflecting the on-PM state after the crash;
\sysname then mounts the target PM file system on the crash state, and checks if it recovers correctly. 
\sysname does not require modifying the file system implementation,
and is compatible with all PM file systems that implement the POSIX interface.
To choose workloads to test, we couple \sysname with a modified version of the ACE~\cite{Mohan18} workload generator and the Syzkaller~\cite{syzkaller} gray-box fuzzer.
While ACE systematically generates small workloads for \sysname to test, Syzkaller generates random workloads driven by code coverage. 

Second, this paper presents an analysis of \totalbugs unique bugs found by \sysname across four PM file systems (\sref{sec-analysis}).
We have reported all \totalbugs bugs upstream;
all except the two bugs in PMFS (which is not actively maintained) have been confirmed by their developers,
and \fixedbugs have already been fixed.
The bugs have severe consequences such as breaking the atomicity of the \texttt{rename()} system call that applications depend on for atomic updates~\cite{ThanuPillaiEtAl14-OSDI}.  
To the best of our knowledge, this is the largest published corpus of PM file-system crash-consistency bugs;
analyzing it yields useful insights for both PM file-system design and efficient crash-consistency testing of PM file systems.
For example, we observe that many bugs are logic or design issues in performance optimizations rather than the missing flush/fence bugs that many PM bug-finding tools target~\cite{Liu2019, Liu2020, pmemcheck, Liu2021, Di2021, Neal20, Fu2021, Gojiara2021}.

\vheading{Why is testing crash-consistency for PM file systems hard?}
The unique architecture of PM file systems creates two challenges for effective testing.
The first challenge is to intercept writes to the storage media.
Existing crash-consistency testing tools such as CrashMonkey~\cite{Mohan18} intercept writes at the block layer,
but PM file systems remove this natural interception point.
Instead, they write directly to the media using processor store instructions,
which would be prohibitively expensive to instrument
and require reasoning about the CPU cache hierarchy and store reordering
to model faithfully.
The second challenge is choosing which crash states to test.
CrashMonkey and Hydra~\cite{Kim2019} only simulate crashes \emph{after} system calls have returned,
as most file systems only make strong crash-consistency guarantees
after \fsync or \vtt{sync()} calls.
PM file systems take advantage of the low-latency, fine-grained media
to offer stronger consistency guarantees
by making \emph{every} file-system operation synchronous and durable.
This means that to find crash consistency bugs,
we must test crashes \emph{in the middle} of system calls.
The fine granularity of PM writes (8 bytes) also leads to more possible crash states than traditional file systems with block-sized writes.

\vheading{\sysname}.
\sysname tackles these challenges by exploiting insights about the way PM file systems are built.
To intercept writes, we observe that the PM file systems we studied all have \emph{centralized persistence functions} that perform writes to PM, rather than using inline assembly at different points in the code. 
For example, there is a \vtt{memcpy()} implementation that uses non-temporal stores to write to PM.
\sysname uses Kprobes~\cite{kprobes} to intercept these functions and record write IOs without modifying the file system.
Intercepting writes at the function level rather than the instruction level greatly reduces overhead.
We further reduce this overhead by noting that only data that is explicitly flushed can be used to provide crash-consistency guarantees.
Rather than intercepting all writes, \sysname only tracks writes that are explicitly flushed or are written with non-temporal stores.

To choose which crash states to test,
we build on this higher-level interception
as well as empirical results about the write patterns of PM file systems.
Intercepting at the function level allows us to see an entire file-system-level write at once,
and so we can coalesce individual 8-byte stores when appropriate;
for example, we need not test each 8-byte write of a 1\,MB \vtt{write()} call separately.
We also observe that the set of \emph{in-flight writes} (writes in volatile caches that have not yet been flushed) at any point in a system call is typically small,
reducing the number of crash states \sysname must explore.
Finally, we observe that the order in which non-overlapping in-flight writes are written to PM does not matter,
and so \sysname only considers different subsets of in-flight writes being persisted rather than all permutations.

\vheading{Generating workloads}.
Given a workload, \sysname provides a mechanism to generate and test crash states;
an orthogonal question is deciding which workloads to explore.
CrashMonkey~\cite{Mohan18} hypothesized that systematically exploring small workloads on small file system states was effective in finding crash-consistency bugs;
we sought to test this hypothesis for PM file systems. 
We modify the Automatic Crash Explorer (ACE) workload generator from CrashMonkey to account for the fact that system calls are synchronous in the tested systems, eliminating the need for \fsync.
To try to invalidate the hypothesis, we also use the Syzkaller~\cite{syzkaller} gray-box kernel fuzzer that can generate longer, more complex workloads. 
We modify Syzkaller to only consider file-system-related system calls and to include recovery code when measuring code coverage.

\vheading{Testing results and observations}.
We use \sysname to test four open-source in-kernel PM file systems: PMFS~\cite{Dulloor2014}, NOVA~\cite{Xu2016}, NOVA-Fortis~\cite{Xu2017}, and WineFS~\cite{Kadekodi21}. 
\sysname finds \totalbugs unique bugs across the four file systems, with two bugs present both in PMFS and WineFS (which builds on PMFS). 
From these results,
we draw some common observations
about PM file systems and how to test them:
\begin{itemize}
\item The CrashMonkey hypothesis about simple workloads finding many crash-consistency bugs also holds for PM file systems: 15 of \totalbugs bugs are found by the systematic exploration of our modified ACE.  

\item A majority of the discovered bugs (11/\totalbugs) could only be found by simulating a crash in the middle of a system call, meaning that existing tools that test only between system calls would not have been effective.

\item While a number of recent tools focus on finding missing or duplicate cache flushes and fences in PM applications~\cite{Liu2019, Liu2020, pmemcheck, Liu2021, Di2021, Neal20, Fu2021, Gojiara2021}, we found that a majority of the bugs (14/18) did not result from such issues, but instead from mistakes in logic (\eg a metadata item that was not added to a transaction).

\item PM file systems increase performance by maintaining some data structures only in DRAM, and rebuilding them when the file system is mounted~\cite{Xu2016, Kadekodi19, Kadekodi21}; \sysname found seven bugs in such code.

\item Six bugs arose from developers trying to increase performance by updating metadata in-place, which is much easier to do with the fine-grained access model of PM, rather than inside a transaction.
  
\item NOVA-Fortis~\cite{Xu2017} contains a number of features not present in NOVA~\cite{Xu2016} that are intended to increase resilience; interestingly, \sysname found five bugs in these complex features.
\end{itemize}  

The analysis contains a number of other observations, along with a discussion of their implications.
To the best of our knowledge, this is the first such analysis of crash-consistency bugs in PM file systems. 

\vspace{5pt} \noindent In summary, this paper makes the following contributions:
\begin{compactitem}
\item A set of tools to test crash-consistency of PM file systems, including the \sysname framework (\sref{sec:tools})
\item A corpus of \totalbugs crash-consistency bugs discovered by these tools across four PM file systems (\sref{sec:testing})
\item An analysis of discovered crash-consistency bugs, with insights for PM file-system design and crash-consistency testing (\sref{sec-analysis})
  \end{compactitem}

\section{Background and Motivation}
\label{sec-bkgd}

This section describes file-system crash consistency.
It then discusses why
crash consistency is important, why testing it for PM file systems is 
challenging, and why existing tools do not solve this problem.

\vheading{Crash consistency}.
A file system is \emph{crash consistent} if it maintains a set of guarantees about its data and metadata after a crash due a power loss or a kernel bug~\cite{ChidambaramPhd15, crash-cacm15, GangerPatt-Metadata94}. 
For example, if there is a crash in the middle of a \vtt{rename()} system call, the POSIX standard requires that the file system after recovery should have the file in either the old name or the new name; in other words, rename must be atomic even if there is a crash \cite{posix2018}.

Many applications depend on the file system to be crash consistent~\cite{ThanuPillaiEtAl14-OSDI}.
Continuing with the rename example, many applications including text editors such as emacs and vim use temporary files to store user data, and rename the temporary files over the original files when the user saves the file.
If rename is not atomic, these applications can lose user data in a crash. 
Unexpected power loss occurs even in professionally-managed data centers~\cite{power-amazon, power-canada, power-dreamhost,
  power-fire, power-internap, power-london}.
Thus, it is important to ensure that file systems are crash consistent. 

\vheading{Persistent memory (PM)}.
Persistent memory technology,
recently commercialized as Intel Optane DC Persistent Memory~\cite{optane, pmmdoc},
combines the properties of traditional storage media and DRAM;
it is byte-addressable and connected to the memory bus like DRAM, but provides persistence like traditional storage media. 
Compared to DRAM, Optane PM provides $0.33\times$ read bandwidth, $0.17\times$ write bandwidth, 2--4$\times$ higher read latency, and similar write latency.

In the x86 programming model,
PM is accessed via processor load and store instructions.
Writes to PM flow through the CPU cache hierarchy like any other memory store,
and so do not become immediately persistent.
Data can be flushed from CPU caches to persistent media with cache line flush instructions (\vtt{clfush}, \vtt{clflushopt}, \vtt{clwb}),
or can bypass cache entirely
with non-temporal stores (\vtt{movnt}).
Because writes to PM are processor stores,
they are also subject to CPU store reordering,
and so must be surrounded by store fences
when preserving order is important for consistency.
We say that data whose cache line has been written back, or which was written using non-temporal stores, is \emph{flushed} to PM once a subsequent store fence instruction has executed,
as it is guaranteed to reach media before any future writes.
We term a write that has not yet been flushed to persistent media an \emph{in-flight} write;
in-flight writes may be lost in the case of a crash.
If there are multiple in-flight writes, they may be written to persistent media in any order.

\vheading{PM file systems}. 
PM file systems differ from traditional file systems in a few important aspects. 
First, traditional file systems have a long software path for writes along the storage stack; all writes go through the block layer, and most through the page cache, before hitting the storage media. 
In the case of PM, writes are performed using processor stores, providing a significantly shorter path from file system to storage media. 
Second, while traditional file systems buffer updates in memory before writing them to storage, PM file systems tend to synchronously write to persistent media given the low latency and high bandwidth of PM~\cite{Kadekodi19, Kadekodi21, Kwon17, Dulloor2014}.
As a result, every system call leads to persistent writes, not just \vtt{fsync()} and \vtt{sync()}.

\noindent These design decisions lead to two challenges when testing PM file systems for crash consistency:
\begin{itemize}
\item \viheading{Intercepting writes is challenging}.
Existing tools to test crash-consistency intercept writes at the block layer and use the intercepted write I/O to construct crash states.
PM file systems remove this natural interception point.
PM file systems write to the persistent media using processor store instructions;
the small granularity of the store instruction (8 bytes) also increases the number of instructions that would need to be intercepted.
A naive solution such as intercepting each store instruction (even if we could correctly identify stores to PM) would be prohibitively expensive.

\item \viheading{Increase in crash states to explore}.
  While existing crash-consistency testing frameworks only simulate crashes at system-call boundaries,
  testing PM file systems requires simulating crashes \emph{in the middle} of system calls.
  This is because PM file systems persist state synchronously in file-system operations, instead of buffering them until \fsync.
  The small granularity of writes also increases the number of potential crash states resulting from each write or metadata operation.
  Testing crash consistency in PM file systems requires simulating crashes in the middle of each file-system operation, versus just after \fsync or \vtt{sync()} calls.
\end{itemize}

\vheading{Why current tools are not enough}.
CrashMonkey~\cite{Mohan18} and Hydra~\cite{Kim2019} can be used to check the crash consistency of traditional file systems.
However, they only simulate crashes at the end of system calls such as \fsync because that is when persistence is guaranteed.
However, PM file systems have stronger persistence guarantees, and we need to simulate crashes in the \emph{middle} of the system call, not only at the end.
It is not straightforward to do this since the persistence guarantees are unclear when crashing in the middle of a system call.
Moreover, CrashMonkey and Hydra cannot intercept PM writes.

Yat~\cite{Lantz2014} is an internal tool used by Intel to test the crash consistency of PM software, including PM file systems.
Yat uses a hypervisor-based recording technique to log all writes to PM and reconstructs crash states to check.
It produces an extremely large number of crash states; one of the three reported test workloads used with PMFS would take over five years to run fully.
Yat is not open source, and the exact set of bugs it found is not available.

A number of tools are available for checking PM software for \emph{PM programming errors}: errors associated with persisting data to PM, such as missing or unnecessary cache-line flushes or store fences~\cite{Liu2019, Liu2020, pmemcheck, Liu2021, Di2021, Neal20, Fu2021, Gojiara2021}.
PMTest \cite{Liu2019}, XFDetector \cite{Liu2020}, Pmemcheck \cite{pmemcheck}, and PMDebugger \cite{Di2021} require manual annotation, which we seek to avoid.
All of them are primarily focused on low-level PM programming errors or narrow classes of logic bugs that are directly related to PM management. For example, Witcher \cite{Fu2021} looks for fine-grained ``persistence atomicity violations'' in which sequences of writes to PM that are assumed to be atomic can be interrupted by a crash. Agamotto \cite{Neal20}, which uses symbolic execution to track the state of PM, also allows developers to provide custom bug oracles to check for precise, low-level properties about a program; for example, that a specific type of structure is always modified within a transaction.
We are interested in checking the higher-level crash consistency guarantees provided by PM file systems and the POSIX interface without requiring specifications from developers, so these tools are not sufficient. 


In summary, there is a strong need for a tool that can be used to efficiently test PM file systems for crash consistency without requiring manual annotations.

\section{\sysname}
\label{sec:tools}

We present \sysname, a new framework
to find crash-consistency bugs in PM file systems.
\sysname tackles the challenges outlined in \sref{sec-bkgd} by exploiting two characteristics of PM file systems: centralized persistence functions and limited number of in-flight writes. 
\sysname can be run on all PM file systems implementing the POSIX API, and it
neither requires manual annotations nor modifying the file-system code. 

\sysname takes as input a workload to execute against a chosen file system
and outputs bug reports with enough details to reproduce the bug. 
We have developed two workload generators:
one based on the ACE tool~\cite{Mohan18} for exhaustively enumerating file-system operations,
and one based on the Syzkaller gray-box fuzzer~\cite{syzkaller}.
Together, these tools offer an automated and off-the-shelf approach
for discovering bugs in PM file systems.

This section first provides an overview of the end-to-end bug finding process (\sref{sec-overview}).
It then describes how \sysname tackles the main challenges in testing PM crash consistency (\sref{sec-challenges}).
It then presents the architecture of \sysname (\sref{sec-architecture})
and  workload generators (\sref{sec:workloads}), followed by a discussion of the limitations of the framework (\sref{sec:limitations}).

\begin{figure}[t]
    \centering
    \includegraphics[width=0.48\textwidth]{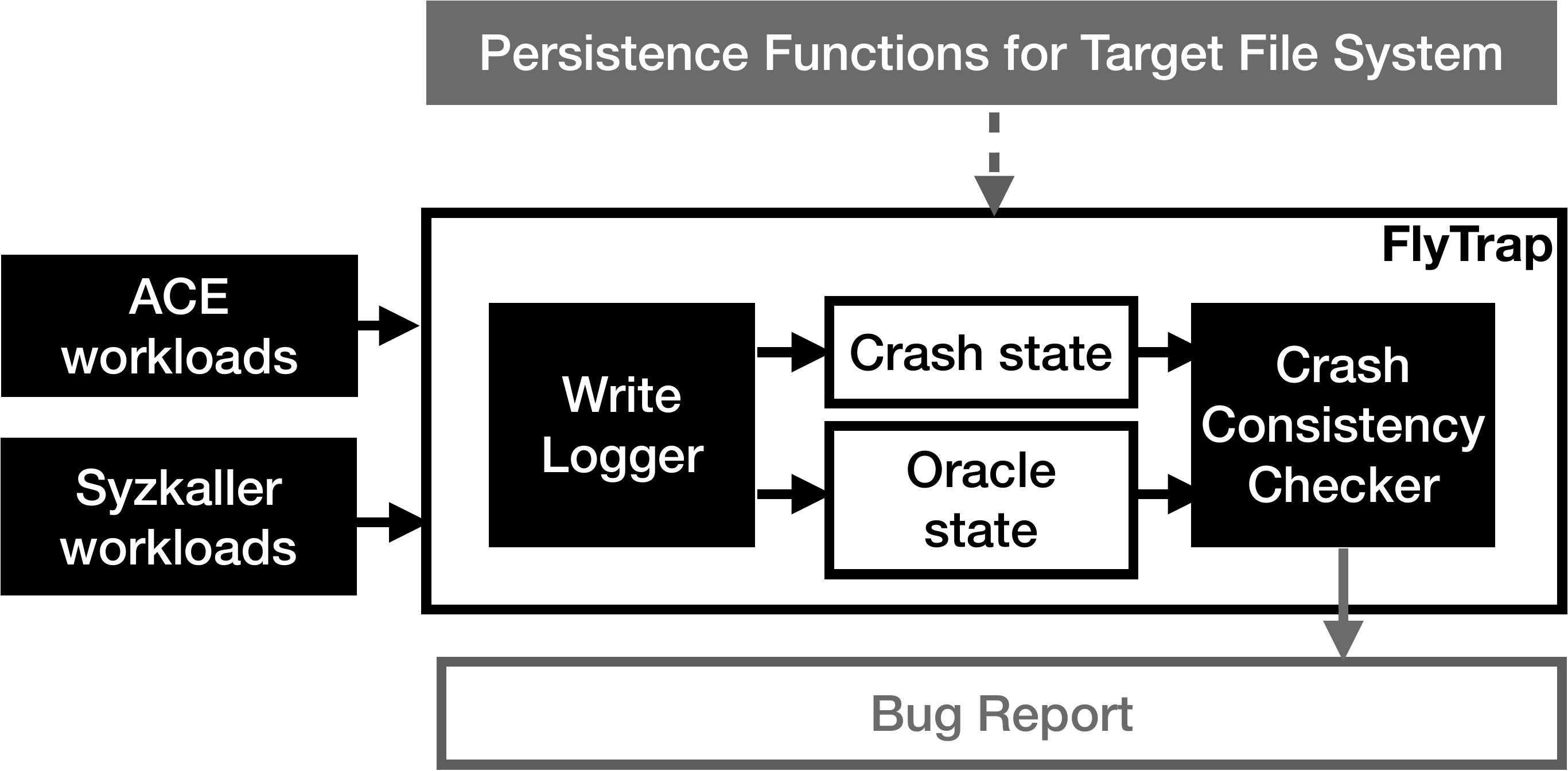}
    \mycaption{Architecture}{Given a target file system and its persistence functions,
      \sysname uses workloads from both ACE and Syzkaller to test the file system.
      \sysname produces bug reports with enough detail to reproduce the bug.
    }
    \label{fig-arch}
\end{figure}

\subsection{Overview}
\label{sec-overview}

Figure~\ref{fig-arch} provides an overview of the \sysname framework. 
\sysname is a record and replay framework.
It first runs a given workload (a sequence of file-system operations) and records the writes made by the file system.
It then replays these writes to create crash images, which represent the state of the system if it had crashed at different points during the workload.
\sysname constructs crash images for crash points both during and after system calls.
\sysname mounts the target file system on the crash image, lets it recover, and then checks whether it has recovered to a consistent state. 
Consistency is workload-specific;
for example, if the crash happened in the middle of a rename, \sysname will check whether the old file or the new file exists.
Both files missing or both files existing would result in a bug report.
The bug report contains the workload, the crash point, the file-system state after crash, what was expected, and the file system and kernel version. 

The ACE workload generator is based on a hypothesis outlined in the CrashMonkey work~\cite{Mohan18}: that testing small workloads (with a few file-system operations) on a newly created file system is useful for finding crash-consistency bugs.
We use the ACE workload generator with \sysname to test if this hypothesis holds for PM file systems.
We modify ACE to take into account the properties of PM file systems, such as all operations being synchronous.

To try to disprove the hypothesis, we also use the \ccfuzzer gray-box fuzzer to generate long, complex, and randomized workloads for \sysname.
Syzkaller tries to generate random tests that increase code coverage.
We modified Syzkaller to take the recovery code into account when calculating code coverage. 

Finally, we cluster bug reports based on their text, such as the error (\eg missing file) and the workload.
This allows us to efficiently find and report unique bugs to developers.

\subsection{Challenges}
\label{sec-challenges}

\sysname tackles the challenges in testing PM file systems for crash consistency (\sref{sec-bkgd}) by exploiting two characteristics of PM file systems. 

\vheading{Intercepting writes}.
Intercepting the write traffic of a PM file system is more complex than  a traditional file system because
PM file systems do not use the usual block layer to write data to the storage media, removing a convenient interception point. Instead,
PM file systems use processor store, fence, and cache flush instructions directly.
However, 
all  in-kernel PM file system implementations we examined include \textit{centralized persistence functions} to interact with the PM device rather than directly using inline assembly at each write location.
In particular, every file system we evaluate in this paper
offered four persistence functions: non-temporal {\tt memcpy}, non-temporal {\tt memset}, cache line flush, and store fence.
These abstractions simplify reasoning about PM semantics and potentially enable portability to other architectures.

We exploit this observation to simplify intercepting writes
by requiring the file system developer
to provide \sysname with annotations
to identifies these four persistence functions for their file system.
\sysname then automatically instruments these functions at run time
using the Kprobes~\cite{kprobes} debugging mechanism in the Linux kernel.

\vheading{Increased number of crash states}.
PM file systems make very fine-grained writes to storage media. Also, since these systems perform operations synchronously, they provide strong guarantees about crash consistency at all times, not just after \fsync. 
As a result, a workload can result in significantly more crash states that are interesting to test in a PM file system than in a block-based file system.
We handle the large number of crash states by exploiting three observations:
\begin{itemize}
\item \emph{Data required for crash consistency must be flushed}.
  While the PM file system may write a lot of data, only data that is explicitly flushed or is written via non-temporal stores can be relied  upon to be persistent.
  CrashMonkey~\cite{Mohan18} and Hydra~\cite{Kim2019} leverage a similar observation to only simulate crashes after \sysfsync and \vtt{sync()}.
  \sysname exploits this observation to only track writes when they are flushed --- it collects write information only for cache flushes and non-temporal writes.
  This drastically reduces the overhead of logging and the number of functions \sysname must track.
\item \emph{Order of in-flight writes does not matter}.
  If there are two in-flight writes $A$ and $B$ (data in volatile CPU caches that has not been flushed yet), the order in which A and B are persisted does not matter.
  A crash state where $A$ is written first followed by $B$ is equivalent to a crash state where $B$ is followed by $A$.
  If $A$ and $B$ are writes to the same cache line, x86 enforces sequential ordering.
  As a result, we do not have to consider re-orderings of in-flight writes.
  \sysname creates crash states using different subsets of in-flight writes, taking into account writes to the same cache line.
\item \emph{Small number of in-flight writes}.
  The number of generated crash states  depends on the number of in-flight writes.
  In our testing, we observed that the number of in-flight writes at any given point of time is small:
  across the two workload generators (Section~\ref{sec:workloads}) and four file systems, 
  the average number of the in-flight writes is three, and the maximum is ten.
  The only exception was the \vtt{write()} system call in PMFS, which we observe to have up to 20 in-flight writes when writing a large amount of data.
  This allows \sysname to exhaustively enumerate and check all subsets of in-flight writes. 
  
\end{itemize}  

Taken together, these observations allow \sysname to track only flushed writes (leading to low overhead), consider only subsets (and not permutations), and exhaustively test all the subsets resulting from in-flight writes. 

\begin{figure}[t]
    \centering
    \includegraphics[width=0.4\textwidth]{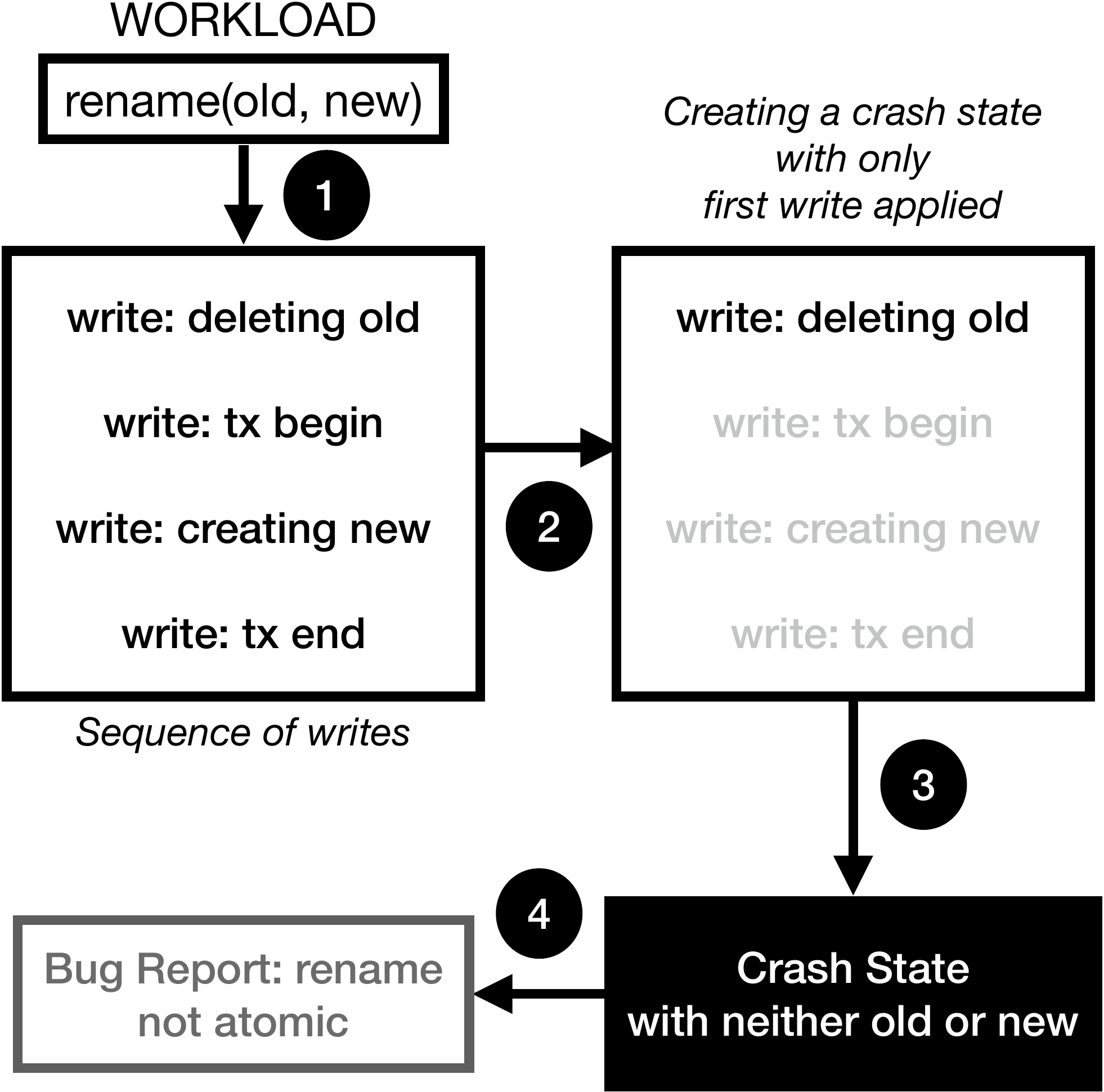}
    \mycaption{\sysname workflow}{
      The figure shows how crash consistency is tested using a simple \vtt{rename()} workload.
      In this example, the old file being deleted is updated in-place, while the new file creation happens inside a transaction.
      1) \sysname runs the workload on the target file system, and logs a sequence of PM writes, flushes, and fences.
      For the sake of simplicity, assume that all writes are flushed and there is a store fence at the end of the system call.
      2) \sysname creates a crash state where only the old file is deleted; the other writes are lost in the crash.
      3) The consistency checker finds that both the old file and the new file are missing
      4) \sysname creates a bug report.
      This bug in NOVA discovered by \sysname (bug \ref{bug:nova4}).
    }
    \label{fig-replay}
\end{figure}

\subsection{\sysname Architecture}
\label{sec-architecture}

\sysname is built on top of the CrashMonkey framework~\cite{Mohan18}, which consists of two kernel modules and user-space utilities to facilitate checking crash states.
We adapt CrashMonkey's user-space utilities to target PM file systems;
the two kernel modules are specific to block devices and are not compatible with PM.
We replace them with modules based on Kprobes, a Linux kernel debugging utility, to record writes to PM.

Given a workload and a target file system, \sysname proceeds in three steps (Figure~\ref{fig-replay}): (1) run the workload and log the writes made by the file system; (2) construct crash states; (3) check each crash state and generate a bug report if required.
We now describe each step in more detail. 

\vheading{Logging writes.}
Recall from Section~\ref{sec-challenges} that \sysname requires PM file system developers to
identify the centralized persistence functions.
Given the locations of these functions, \sysname uses Kprobes~\cite{kprobes} to automatically instrument them at run time. 
Each time one of these functions is invoked by the file system,
the  instrumentation logs the operation and its arguments
(including the current value of the cache line for cache line flushes).
The user-space test harness also inserts markers into this log
to record the start and end of each system call.
This approach requires no code changes to the file-system implementation
other than to prevent the compiler from inlining the persistence functions.
In our experience, identifying these functions was simple,
and we expect it to be even simpler for file-system developers
who likely already know where the persistence functions are in their implementation.

\vheading{Constructing crash states.}
Given a workload, \sysname can simulate crashes both after and during system calls. \sysname replays a workload by walking through the log of writes. Whenever it encounters a cache line write back or non-temporal store, it adds it to an in-flight vector. When it encounters a store fence, it flushes the contents of the in-flight vector. To generate crash states after system calls, \sysname replays all writes that have been flushed by the end of the system call. To generate crash states during system calls, it simulates a crash immediately before each store fence and creates a set of crash states by replaying each subset of the in-flight vector.
The number of crash states is thus dependent on the number of in-flight writes; if there are $n$ in-flight writes, there will be $2^n -1$ crash states.
As noted in Section~\ref{sec-challenges},
we have observed that $n$ is small in practice, allowing \sysname to apply this exhaustive testing strategy. Since a small number of in-flight writes is not a guarantee and we occasionally see larger sets while using Syzkaller, \sysname can place a configurable cap on the number of writes to replay. We find that in practice, even a cap of two writes is sufficient to reveal many bugs (\sref{sec:observations}).

\vheading{Testing each crash state.}
To check file-system consistency,
\sysname first mounts the target file system on each crash state,
which is itself a useful consistency check as failure to mount is a serious bug.
Once successfully mounted, the file-system state is compared against
two oracle file-system states:
a \emph{before} state representing the file-system state before the crashing system call began
and an \emph{after} state representing the file-system state if the system call during the crash successfully completed.
After recovery, the file-system state should match either the before or after state.
This check validates properties implied by POSIX
or widely expected by users in practice~\cite{bornholt:ferrite, ThanuPillaiEtAl14-OSDI}.
Finally, to ensure the file system is in a usable state,
\sysname creates files in all directories and then deletes all files.
If any of these checks fail,
the checker reports a bug describing the inconsistency
and the corresponding crash state.

To make this implementation efficient,
we construct oracle file systems incrementally
as the log is replayed because many crash states share the same oracle state.
We also construct crash states in-place on the PM device,
mutating them to generate each crash state
rather than recreating them from scratch each time.
Because the consistency checks mutate the state,
we reuse our logging infrastructure
to record an undo log for these mutations
and roll back the changes when advancing to the next crash state.

\subsection{Workload Generation}
\label{sec:workloads}

Given a workload, \sysname generates crash states and tests them for consistency.
An orthogonal challenge is generating workloads for \sysname to test.
The CrashMonkey work~\cite{Mohan18} introduced the hypothesis that small workloads on new file systems are useful in finding crash-consistency bugs.
While this hypothesis was true on traditional block-based file systems, we aim to test whether it holds on PM file systems.
To this end, we modify CrashMonkey's Automated Crash Explorer (ACE), which systematically explores workloads of a given size, to work with \sysname.
To disprove the hypothesis, we modify the Syzkaller~\cite{syzkaller} gray-box fuzzer to work with \sysname.
Syzkaller generates long, complex, randomized workloads while aiming to improve code coverage.

\subsubsection{Automatic Crash Explorer}

We used a modified version of ACE \cite{Mohan18} to systematically generate workloads for \sysname.
ACE was designed to exhaustively generate workloads of a certain structure to test traditional file systems.
It focuses on short workloads with frequent {\tt fsync()}, {\tt fdatasync()}, or {\tt sync()} calls, as these are required to obtain strong crash-consistency guarantees in traditional file systems.
Given a sequence length $n$, ACE generates workloads with $n$ core file-system operations over a small, predetermined set of files, then fleshes them out by satisfying dependencies and adding {\tt fsync()}, {\tt fdatasync()}, or {\tt sync()} operations.
A workload with $n$ core system calls is called a ``seq-$n$" workload.

We modify ACE in the following manner for \sysname.
First, since system calls in the {\tt fsync} family do not have the same crash-consistency significance in PM file systems as they do in traditional file systems, we remove them from our version of ACE.
This significantly reduces the number of workloads generated because we do not need to consider workloads that differ only by the sync calls they use.
We also remove system calls and options that are not supported by the file systems under test.
For example, NOVA only supports the {\tt FALLOC\_FL\_KEEP\_SIZE} flag for the {\tt falloc} system call, and it does not support extended attributes.
Removing these unsupported options further reduces the number of workloads that ACE generates.

We test all seq-1 and seq-2 workloads, as well as the subset of seq-3 workloads
containing only {\tt pwrite()}, {\tt link()}, {\tt unlink()}, and {\tt rename()} calls (i.e. the ``seq-3 metadata'' workloads in the CrashMonkey work\cite{Mohan18}) to make testing tractable.
Our modified version of ACE generates 56 seq-1 tests, 3136 seq-2 tests, and 50650 seq-3 metadata tests.

\subsubsection{Syzkaller}

We modify Syzkaller~\cite{syzkaller}, a state-of-the-art gray-box kernel fuzzer, to generate workloads for \sysname.
As is standard in gray-box fuzzing, our fuzzer starts with an initial set of test cases (seeds) and uses genetic programming to generate new tests for \sysname from those seeds.
\sysname tests each generated workload and reports back whether the workload produced a crash-consistency bug along with the code coverage achieved on the target kernel.
If the workload covered new parts of the kernel, the fuzzer adds it to its set of seeds and generates new workloads from it.

Syzkaller generates workloads by randomly selecting sequences of core file-system operations and their argument values.
It generates syntactically and semantically valid workloads by using a detailed template for each system call
that specifies more precise \emph{qualified type} information~\cite{Millstein2005}
for the call's arguments.
For example, the template for {\tt open} specifies that it returns a file descriptor, not just an arbitrary integer;
the template for {\tt write} specifies that its first argument is also a file descriptor rather than an arbitrary integer
(which would be very unlikely to be a valid file descriptor). 

To adapt Syzkaller to our setting,
we restrict it to only generate workloads that contain core file-system operations,
and replace its workload executor with a custom one.
Our executor invokes \sysname on each workload
and records code coverage both before the crash and during recovery.
We add a workload to the seed set if it achieves new code coverage
on either side of the crash.

Like many fuzzers, \ccfuzzer can quickly generate many bug reports that are duplicates---we found that \ccfuzzer would frequently generate different workloads that triggered the same bug and produced similar bug reports. In our setting, this duplication also arises when multiple crash states for the same workload trigger the same bug, producing similar bug reports.
To address this problem, we extended Syzkaller to automatically triage bug reports generated by \sysname during fuzzing.
We use a simple triaging procedure that clusters bug reports by lexical similarity. Whenever \sysname generates a bug report, we compute the report's word vector $v$ (based on its text),
and then compute the distance from $v$ to the closest cluster $C$ of previous reports.
If the distance from $v$ to $C$ is below a predefined threshold, we add the report to $C$,
otherwise we create a new cluster.
This simple heuristic was more effective than more complex ones we tested.
We also updated Syzkaller to display these bug report clusters (along with the workload that triggered each bug report) in its UI dashboard to make them easier for users to debug.

\subsection{Limitations}
\label{sec:limitations}

\sysname has several limitations.
First, it can miss bugs
because it does not explore all workloads
nor all crash states resulting from a given workload.
\sysname only tests some subsets of the data in \vtt{write()} system calls and considers cache line flushes atomic with respect to crashes.
\sysname only tracks flushed writes; a PM write that is never flushed is invisible to \sysname, and it could result in \sysname missing a bug.
For example, consider file-system writes to $A$, $B$, $C$,
and a bug where the file system never flushed $A$.
\sysname would not explore a crash state where $A$ and $B$ are persisted, but $C$ is not,
and so if this state created inconsistency \sysname would not detect it.
Second, \sysname does not support checking concurrent workloads.
It assumes that the operations performed by the file system in each workload are deterministic.
In our testing so far, we have not observed any non-determinism in the file systems that impacted our ability to find and reproduce bugs with \sysname.
Third, \sysname assumes that the PM file system has centralized persistence functions.
A PM file system that uses in-line assembly to update PM at various locations in its code would not be compatible with \sysname.

Despite these limitations, we believe that \sysname is a useful addition to the set of tools for building robust PM file systems. In particular,
the level of automation provided by \sysname allows developers to test new or in-development PM file systems efficiently. 

\section{Testing PM File Systems}
\label{sec:testing}

In this section, we evaluate \sysname's effectiveness at finding bugs across different PM file systems.
We describe our experimental setup (\sref{sec:setup}) and present an evaluation of \sysname along with a comparison of ACE and Syzkaller as workload generation strategies (\sref{sec:eval}).
We provide a brief overview of the bugs found by \sysname (\sref{sec:results}).

\subsection{Experimental setup} \label{sec:setup}

\paragraph{File systems.}
We considered seven open-source PM file systems for testing with \sysname: NOVA~\cite{Xu2016}, NOVA-Fortis~\cite{Xu2017}, PMFS~\cite{Dulloor2014}, WineFS~\cite{Kadekodi21}, Strata~\cite{Kwon17}, Assise~\cite{Anderson20}, and ext4-DAX~\cite{daxfiles}.
ext4-DAX is a modification of the ext4 file system for PM. It shares most of its code with ext4 and has the same crash consistency guarantees, so tools like CrashMonkey or Hydra that are designed for block-based file systems are a better fit for testing ext4-DAX. Strata and Assise are kernel-bypass file systems implemented in user space. Neither Strata nor Assise currently support recovering from arbitrary crashes. As a result, we chose to focus on the remaining four systems: NOVA, NOVA-Fortis, PMFS, and WineFS.

\paragraph{Test infrastructure.}
All experiments described in this paper were run on QEMU/KVM virtual machines running Debian Stretch. Each VM is allocated one CPU (except for those testing WineFS, which requires four CPUs) and 8 GB of RAM. Each VM also has two 128 MB emulated PM devices, which are used to execute the workload, construct the oracle file system, and check crash states.

We run ACE-generated workloads on a single Amazon EC2 m5d.metal instance with 96 vCPUs, 384 GB memory, and four 900 GB NVMe SSDs. We use these resources to check multiple file systems using workloads of multiple sequence lengths in parallel. We run seq-1 and seq-2 tests on individual VMs; we split the seq-3 metadata workloads across 10 VMs and ran them in parallel. All file systems were checked using seq-1 and seq-2 workloads, and all but WineFS were run on seq-3 metadata tests. At the time we ran these experiments, WineFS had a bug that prevented creation of the number of files some seq-3 tests require. The number of in-flight writes at any time during ACE tests is consistently low, so we do not place a cap on the number of crash states for ACE.

To evaluate \sysname with Syzkaller, we run four Chameleon Cloud \cite{Keahey2020} bare metal instances, which have two Intel Xeon Gold 6240R CPUs each with 24 cores and 48 threads, as well as 192 GB RAM, and 480 GB storage. Each host fuzzed a different file system using 15 virtual machines. Each fuzzer starts with an empty set of seeds. Syzkaller-generated tests can be long and generate many crash states, so to avoid the fuzzer getting stuck, we run \sysname with a cap of two writes per crash state; as \sref{sec:obs_testing} observes, this cap does not affect its ability to find bugs in practice.

\subsection{Evaluation} \label{sec:eval}

\begin{figure}
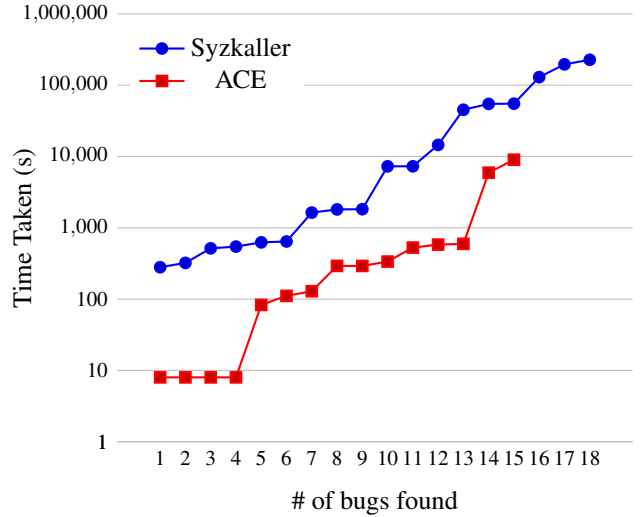

\pgfplotsset{scaled y ticks=false}
\plot{results/TimeToFindBugs.csv}{}
\caption{Cumulative time taken to find crash-consistency bugs by ACE and Syzkaller.} 
\label{fig:bugtiming}
\end{figure}

\vheading{ACE tests.} For each file system under test, \sysname took about 10--15 minutes to run the seq-1 workloads, 7--11 hours to run the seq-2 workloads, 16--26 hours to run the seq-3 metadata workloads in parallel (160--260 total CPU hours). The number of crash states to check on each workload varies as much as 3\myx between file systems, with PMFS generally checking the most and WineFS checking the fewest. Overall, \sysname found 15 bugs using ACE tests across all four file systems.

\vheading{Syzkaller.} We ran \sysname with Syzkaller for 18 hours on 15 VMs, for a total of 270 CPU hours spent fuzzing. During this time, \sysname checked over 40 million crash states across all four system calls, finding 18 unique bugs. Three of these bugs cannot be found with ACE-generated workloads.

\vheading{Comparison.} We ran \ccfuzzer and ACE on each file system and recorded the cumulative CPU time taken to find all bugs when using each workload generator.
Figure~\ref{fig:bugtiming} shows the result of this experiment.
ACE finds the first 15 out of \totalbugs bugs in less than three CPU hours total, but is unable to find the final three bugs.
\ccfuzzer, on the other hand, takes almost 20\myx more CPU time than ACE to find the first 10 bugs and almost 6\myx more CPU time to find all the bugs the ACE tester finds.
However, when we let \ccfuzzer run for an additional 27 CPU hours, it is able to find three additional bugs that are not detected by ACE. ACE misses these three bugs because they do not conform to the patterns that it uses to generate workloads. For example, two of these bugs create two open file descriptors to the same file and modify the file's contents through both file descriptors, but ACE does not generate workloads with multiple file descriptors for the same file.

While the results of this experiment indicate that \ccfuzzer has greater overall bug finding capability than ACE, the ACE tests are considerably more resource efficient.
This suggests that the ACE tests can be run locally and iteratively to find bugs during file-system development,
whereas \ccfuzzer should be run for a long time in an environment with ample compute resources (such as the cloud) for more comprehensive crash-consistency testing.


\begin{table}
    \centering
    \begin{tabular}{lc}
        \toprule
        System call & \# of bugs \\
        \midrule
        \vtt{mkdir} & 1 \\
        \vtt{creat} & 1 \\
        \vtt{rmdir} & 1 \\ 
        \vtt{fallocate} & 1 \\
        \vtt{unlink} & 2 \\
        \vtt{link} & 3 \\
        \vtt{rename} & 4 \\
        \vtt{truncate} & 5 \\ 
        \vtt{write}/\vtt{pwrite} & 5 \\
        \bottomrule
    \end{tabular}
    \caption{The number of bugs associated with each system call. Some bugs impact all file system operations, and some impact multiple system calls.}
    \label{tab:syscall_bug_counts}
\end{table}

\begin{table*}
    \centering 
    \begin{tabular}{llp{.35\textwidth}p{.25\textwidth}l}
        \toprule
        Bug \# & File System & Consequence & Affected system calls & Type \\ \midrule
        \begin{bug} \label{bug:nova1} \end{bug} & NOVA & File system unmountable & All & Logic \\
        \begin{bug} \label{bug:nova2} \end{bug} & NOVA & File is unreadable and undeletable & {\tt mkdir, creat} & PM \\
        \begin{bug} \label{bug:nova3} \end{bug} & NOVA & File system unmountable & {\tt write, pwrite, link, unlink, rename} & Logic \\
        \begin{bug} \label{bug:nova4} \end{bug} & NOVA & Rename atomicity broken (file disappears) & {\tt rename} & Logic \\
        \begin{bug} \label{bug:nova5} \end{bug} & NOVA & Rename atomicity broken (old file still present) & {\tt rename} & Logic \\
        \begin{bug} \label{bug:nova6} \end{bug} & NOVA & Link count incremented before new file appears & {\tt link} & Logic \\ 
        \begin{bug} \label{bug:nova7} \end{bug} & NOVA & File data lost & {\tt truncate} & Logic \\ 
        \begin{bug} \label{bug:nova8} \end{bug} & NOVA & File data lost & {\tt fallocate} & Logic \\
        \begin{bug} \label{bug:nova-fortis1} \end{bug} & NOVA-Fortis & Unreadable directory or file data loss & {\tt unlink, rmdir, truncate} & PM \\
        \begin{bug} \label{bug:nova-fortis2} \end{bug} & NOVA-Fortis & File is undeletable & {\tt write, pwrite, link, rename} & Logic \\
        \begin{bug} \label{bug:nova-fortis3} \end{bug} & NOVA-Fortis & FS attempts to deallocate free blocks & {\tt truncate} & Logic \\ 
        \begin{bug} \label{bug:nova-fortis4} \end{bug} & NOVA-Fortis & File is unreadable & {\tt truncate} & Logic \\
        \begin{bug} \label{bug:pmfs1} \end{bug} & PMFS & File system unmountable & {\tt truncate} & Logic \\
        \begin{bug} \label{bug:pmfs2} \end{bug}\&\begin{bug} \label{bug:wine2} \end{bug} & PMFS\&WineFS & Write is not synchronous & {\tt write, pwrite} & PM \\
        \begin{bug} \label{bug:pmfs3} \end{bug} & PMFS & Out-of-bounds memory access & All & Logic \\
        \begin{bug} \label{bug:pmfs4} \end{bug}\&\begin{bug} \label{bug:wine3} \end{bug} & PMFS\&WineFS & File data lost & {\tt write, pwrite} & PM \\
        \begin{bug} \label{bug:wine1} \end{bug} & WineFS & File is unreadable and undeletable & All & Logic \\
        \begin{bug} \label{bug:wine4} \end{bug} & WineFS & Data write is not atomic in strict mode & {\tt write, pwrite} & Logic \\
        \bottomrule
    \end{tabular}
    \caption{Bugs found by \sysname, their consequences, and the system calls that they affect.}
    \label{tab:bug_description}
\end{table*}

\subsection{Results} \label{sec:results}

\vheading{Crash-consistency bugs.} Using ACE- and \ccfuzzer-generated tests, \sysname finds \uniquebugs unique crash-consistency bugs across all four tested file systems. The number of unique bugs is based on the number of separate fixes required to patch all of the bugs, not different user-visible consequences. Two bugs are found in both WineFS and PMFS (since WineFS is built off PMFS), for a total of 20 bugs. 

Table~\ref{tab:bug_description} describes the consequences of each bug and the system calls they affect. The bugs are classified as either logic or PM errors (described further in \sref{sec:observations}). \sysname found eight bugs in NOVA, four bugs in NOVA-Fortis, two bugs in PMFS, two bugs in WineFS, and two bugs in both PMFS and WineFS (as WineFS is an extension of PMFS). Many of these bugs have serious consequences that violate the crash consistency guarantees of the file systems. Three of the bugs prevent the file system from being mounted entirely. Two impact the atomicity of \vtt{rename()}, which many applications rely on~\cite{ThanuPillaiEtAl14-OSDI}.
Many others cause data loss or prevent a user from accessing files entirely.

The bugs impact a wide variety of system calls, as Table~\ref{tab:syscall_bug_counts} shows. Many bugs are located in common metadata-handling or recovery code and thus impact multiple system calls. \vtt{rename()}, \vtt{write()}/\vtt{pwrite()}, \vtt{link()}, and \vtt{truncate()} are especially bug-prone and are affected by multiple bugs.

\vheading{Non-crash consistency bugs.} While working with \sysname, we also found an additional eight non-crash-consistency bugs
not included in Table~\ref{tab:bug_description}.
These bugs occur during regular execution and do not require a crash to be exposed.
We were able to find these bugs because they caused KASAN errors, segmentation faults, or incorrect behavior that our consistency checks could detect.
For example, using the fuzzer, we discovered that NOVA does not properly handle {\tt write()} calls where the number of bytes to write is extremely large;
it will allocate all remaining space for the file, causing most subsequent operations to fail.
\sysname detected this bug because it found that new files could not be created in crash states following this write.

\section{Bug Analysis}
\label{sec-analysis}

\begin{table*}[t]
    \centering
    \begin{tabular}{p{.73\textwidth}p{.22\textwidth}}
        \toprule
        \textbf{Observation} & \textbf{Associated bugs} \\ \midrule
        Many bugs occur due to logic or design issues, not PM programming errors. & \ref{bug:nova1}, \ref{bug:nova3}--\ref{bug:nova8}, \ref{bug:nova-fortis2}--\ref{bug:pmfs1}, \ref{bug:pmfs3}, \ref{bug:wine1}, \ref{bug:wine4} \\ [0.1cm]
        
        The complexity of performing in-place updates leads to bugs. & \ref{bug:nova4}--\ref{bug:nova7}, \ref{bug:pmfs2}, \ref{bug:wine2} \\ [0.1cm]
        
        Recovery code that deals with rebuilding in-DRAM state is a significant source of bugs. & \ref{bug:nova1}, \ref{bug:nova3}, \ref{bug:nova7}, \ref{bug:nova-fortis3}, \ref{bug:pmfs1}, \ref{bug:pmfs3}, \ref{bug:wine1} \\ [0.1cm]
        
        Complex new features for increasing resilience can introduce new crash consistency bugs. & \ref{bug:nova2}, \ref{bug:nova-fortis1}--\ref{bug:nova-fortis4} \\ [0.1cm]
        
        A majority of observed bugs can only be exposed by simulating crashes during system calls. & \ref{bug:nova3}--\ref{bug:nova6}, \ref{bug:nova-fortis1}--\ref{bug:pmfs1}, \ref{bug:wine1}, \ref{bug:wine4} \\ [0.1cm]
        
        Short workloads were sufficient to expose many crash consistency bugs. & \ref{bug:nova1}--\ref{bug:nova6}, \ref{bug:nova-fortis1}--\ref{bug:wine4} \\[0.1cm]
        
        Many bugs were exposed by replaying a few small writes onto previously persistent state. & \ref{bug:nova3}--\ref{bug:nova6}, \ref{bug:nova-fortis1}--\ref{bug:pmfs1}, \ref{bug:wine1}, \ref{bug:wine4} \\ 
        \bottomrule
    \end{tabular}
    \caption{Observations and the bugs associated with them.}
    \label{tab:observations}
\end{table*}

This sections presents an analysis of the \totalbugs crash-consistency bugs found by \sysname (Table~\ref{tab:bug_description}).
To the best of our knowledge, this is the first such corpus of crash-consistency bugs in PM file systems.
We make some observations about common patterns we found among the bugs (\sref{sec:observations}),
and distill lessons for building and testing PM file systems (\sref{sec:lessons}).


\subsection{Observations}\label{sec:observations}

We first present observations about the nature of the crash-consistency bugs found by \sysname,
and then present observations about crash-consistency testing. 

\subsubsection{Nature of crash-consistency bugs}

\begin{observation}[A majority of the observed crash-consistency bugs are logic issues rather than PM programming errors.]\label{obs:logic}
When we began working on \sysname,
we expected that most bugs would be due to the subtleties
of the persistent-memory programming model
that existing PM application research has focused on:
interaction with the cache hierarchy,
CPU store (re)ordering, etc.
However, the majority of bugs we found---%
14 of \uniquebugs---%
are actually due to logic bugs in manipulating file-system data structures
rather than mistakes in managing persistent memory.
The ``type'' column in Table~\ref{tab:bug_description}
classifies bugs into logic bugs or PM bugs.
Logic bugs are issues that cannot be fixed by adding cache line flushes or store fences.
These results suggest that it is not sufficient for a file-system crash-consistency testing tool
to focus on exploring the persistence behavior of individual writes and reorderings;
it must also exercise the file-system data structures and recovery mechanisms,
and check higher-level consistency properties
that cannot be validated at the level of individual writes.
\end{observation}

\begin{observation}[In-place update optimizations are a common source of crash consistency bugs.] \label{obs:inplace}
One of the allures of persistent memory
is that programs can access it as memory,
performing fine-grained reads and writes directly
rather than coalescing them into larger block-sized IO operations on slower storage media.
This design makes it possible in principle
to reduce the overheads of traditional consistency mechanisms like journaling
by manipulating on-disk data structures directly.
All four file systems we tested use a journal for consistency,
but all have performance optimizations to bypass the journal
and perform in-place updates in certain circumstances.
For example, NOVA updates the link count of a file by updating a per-inode log.
Appending to this log is done via a journalled transaction,
but if the previous operation on the file also updated its link count,
NOVA may modify that log entry in place instead.

We found these optimizations to be particularly error-prone:
6 of \uniquebugs bugs in Table~\ref{tab:bug_description}
are caused by in-place updates.
For example, in bug~\ref{bug:nova4}, NOVA's {\tt rename} implementation
removes the directory entry from the parent inode in-place
but journals the other metadata changes,
allowing the file to be lost in a crash before the journal transaction commits.

Fixing these bugs often requires journalling more data, which is not free.
To quantify the impact of fixing such bugs,
we compared the performance of NOVA before and after
fixing two rename atomicity bugs (\ref{bug:nova4} and \ref{bug:nova5})
by journalling more metadata.
We tested both versions on Intel Optane DC Persistent Memory media.
In a microbenchmark that repeatedly
creates and renames a file over the top of an existing file,
the fixed version of NOVA is 25\% slower.
A more real-world metadata-intensive benchmark
(checking out different stable versions in the Linux kernel git repository)
shows negligible overhead ($<$1\%).
In some cases, journalling can even be better than in-place updates.
The fix for bug~\ref{bug:nova6} replaces an in-place update in {\tt link}
with extra logging,
but makes a microbenchmark that repeatedly creates links to a file 7\% faster,
likely because checking whether the in-place update is safe
requires an extra read from the media.

\end{observation}

\begin{observation}[Rebuilding volatile state during crash recovery is error-prone.] \label{obs:recovery}
In a traditional file system, crash recovery scans on-disk structures
like journals and updates the durable state to match.
PM file systems often perform more work during recovery.
As a performance and write-endurance optimization,
they avoid durably storing some metadata
(e.g., free page lists or file sizes)
and instead rebuild it in volatile memory at mount time.
This rebuilding code is subtle
because it must account for potential inconsistencies or partial states after a crash,
and we found that 8 of the \uniquebugs bugs in Table~\ref{tab:bug_description}
were in such code.
For example, bug~\ref{bug:pmfs1} is caused by a crash during a {\tt truncate} system call on PMFS.
The implementation first stores information about the operation in a ``truncate list'';
if the system crashes before the truncation is complete,
the truncate list can be replayed to finish the operation.
However, replaying truncations requires accessing the free page list,
which PMFS rebuilds in volatile memory during recovery,
but only \emph{after} replaying the truncate list.
Attempts to replay truncations therefore cause a null pointer dereference when accessing the free page list.

Rebuilding volatile state is more complex
in PM file systems that maintain \emph{per-CPU} volatile state
to improve scalability.
In bug \ref{bug:wine1},
WineFS failed to properly index into an array of per-CPU journals
that were read during crash recovery.
\sysname can even find bugs in rebuilding code in the absence of crashes,
because the crash states it tests include ones where \emph{all} writes are durable.
For example, bug~\ref{bug:nova7} involves modifying the same file
through multiple file descriptors on NOVA,
which creates log entries out of chronological order,
violating assumptions in the rebuilding code
and causing data loss during {\tt truncate} operations.

\end{observation}

\begin{observation}[Resilience mechanisms to recover from media failures can introduce new crash-consistency bugs.] \label{obs:fortis}
NOVA-Fortis~\cite{Xu2017} is an extension of NOVA
that adds fault detection and tolerance
for media errors and software bugs
by (among other techniques)
replicating and checksumming inodes and logs
and checksumming file data.
While NOVA-Fortis is not explicitly designed to increase crash resilience,
we tested these features to see if it is more tolerant of crashes than the original NOVA.

We found that NOVA-Fortis has all the same crash-consistency bugs we found in the original version of NOVA,
and in addition has five new bugs caused by the added complexity
of maintaining redundant state and checksums.
A common theme in these bugs
is that data and metadata modifications
are often not atomic
with checksum and replica updates,
allowing checksum validation to fail
(and render a file inaccessible)
even if the file system is consistent and data intact.

\end{observation}

\subsubsection{Crash-consistency testing in PM file systems} \label{sec:obs_testing}

\begin{observation}[A majority of observed bugs require simulating crashes during system calls.] \label{obs:syscall}
Current crash-consistency testing tools for traditional file systems,
like CrashMonkey~\cite{Mohan18} and Hydra~\cite{Kim2019},
insert crashes only at the end of persistence system calls ({\tt fsync}, {\tt fdatasync}, etc.).
This heuristic exploits the fact that most POSIX APIs
only make crash-consistency guarantees after persistence operations,
so intermediate states are unlikely to violate the specification.
It allows these tools to scale to test larger workloads,
and does not appear to cause them to miss bugs:
CrashMonkey has a mode to insert crashes during system calls,
but it did not find any additional bugs
in the file systems tested in that work.

We found that this same heuristic does not work for PM file systems.
11 of the \uniquebugs bugs in Table~\ref{tab:bug_description}
require a crash to occur during a system call.
This is a corollary of our observation that most PM file systems
intend to implement most system calls synchronously,
and so their effects are fully persistent by the end of the system call.
For example, the rename atomicity bugs in NOVA (bugs~\ref{bug:nova4} and \ref{bug:nova5})
arise when a crash during the system call leaves only some writes persisted.
Waiting until a persistence point to check consistency would not discover these bugs,
as NOVA correctly flushes all writes by the end of the rename operation.


\end{observation}

\begin{observation}[Short workloads suffice to expose many crash-consistency bugs.] \label{obs:workload}
We use ACE~\cite{Mohan18} to systematically generate test workloads.
ACE's design focuses on exhaustively enumerating small workloads,
based on an empirical study of historical crash-consistency bugs in block-based file systems that showed
most could be reproduced with at most three operations.
It was unclear whether this would hold for PM file systems. 
However, 15 of the \uniquebugs bugs we found in PM file systems
can be found using ACE with at most three operations,
suggesting that this same \emph{small-scope hypothesis}~\cite{jackson:small-scope}
holds for PM file systems.
To try to invalidate this hypothesis,
we also run \sysname using the Syzkaller gray-box fuzzer,
which can generate much longer workloads
but without the exhaustiveness guarantees of ACE (\sref{sec:workloads}).
Syzkaller found 3 bugs that ACE did not.
However, all three bugs were found on short workloads:
two would be considered seq-2 and one seq-3
in terms of the number of core system calls required.
ACE missed them not because of size
but because of complexities that ACE omits
to make exhaustive enumeration tractable,
such as testing unaligned writes.


\end{observation}

\begin{observation}[Most of the observed buggy crash states involve few writes.] \label{obs:crashstates}
\sysname generates crash states by snapshotting known-persistent disk states
between store fences,
and then replaying all subsets of the in-flight writes
between each store fence (\sref{sec:tools}).
For a system call with $n$ in-flight writes before a fence,
this means \sysname should consider all $2^n - 1$ possible crash states.
However, we found that most bugs found by \sysname
involve crash states that include \emph{small} subsets of the in-flight writes.
Of the 11 bugs in Table~\ref{tab:bug_description}
that involve a crash in the middle of a system call,
10 can be exposed by a crash state
that replays only a \emph{single} write
onto the last known-persistent state;
the final bug requires two writes.
This observation suggests a profitable heuristic
for rapid crash-consistency testing
would be to only test small subsets of in-flight rights rather than all of them.
\sysname exploits this observation
by enumerating crash states in increasing order of subset size,
allowing it to find most crash-consistency bugs quickly.
In our experiments, we often cap the number of writes 
that are replayed to build each crash state, 
primarily to prevent \ccfuzzer~from spending many hours checking a single outlier test
that has a high in-flight write count. 
The highest in-flight write count we observed, 20 writes in some PMFS \vtt{write()} calls, would take about 30 hours to check exhaustively using \sysname.
A cap of two is enough to find all bugs presented in this paper; 
a cap of five is sufficient to check all crash states 
for most system calls in the PM file systems we tested.
\end{observation}

\subsection{Lessons Learned}\label{sec:lessons}

Based on our observations above,
we have distilled three lessons for developers of PM file systems
and for building the testing tools that support them.

\begin{lesson}[Synchronous crash consistency on PM file systems simplifies the user experience, but complicates implementation and testing.]
Crash-consistency guarantees in modern file systems are something of a vicious cycle.
File-system developers argue that relaxed guarantees are required to extract reasonable performance~\cite{dpkg:ext4},
but these weak guarantees are a pain point for application developers
and have caused severe data loss in popular applications~\cite{corbet:ext4,boichat:ext4,ThanuPillaiEtAl14-OSDI},
so file-system developers implement workarounds to ``fix'' common mistaken application patterns
and make the intended guarantees even less clear.
The fine write granularity and low latency of persistent memory
finally offers a path to strengthen file-system crash-consistency models,
making resilient applications easier to build and validate.
PM file system developers have taken advantage of this opportunity
by making all system calls synchronous and durable.

While this end result is exciting,
implementing it correctly carries new risks for PM file-system developers
compared to traditional file systems.
We found that many PM file-system bugs come from complex optimizations
to realize high-performance synchronous crash-consistency---%
combining in-place updates with other consistency mechanisms,
and replacing persistent state with reconstructible volatile state---%
that are uncommon techniques on slower storage media.
This is a rich new design space for storage systems,
and identifying the right primitives for this optimizations will be good future work.
These optimizations also create complexity for testing and validation of PM file systems,
which we found requires driving the file system into exercising deeper data structure manipulations
and recovery mechanisms than existing crash-consistency tools are capable of.
\end{lesson}

\begin{lesson}[Diverse testing mechanisms and checkers help invalidate assumptions about crash-consistency patterns.]
Most crash-consistency testing tools build on heuristics and patterns in historic bugs
to select the workloads they test.
We expected to bring those patterns across to PM file systems,
focusing on short workloads and a small set of potential crash points and patterns.
However, we found instead that most assumptions about file-system crash consistency
do not carry across to persistent memory,
where the consistency mechanisms and guarantees are significantly different.
Finding crash consistency bugs in PM file systems
requires exploring many more crash states than other file systems,
including crashes in the middle of system calls;
we had to develop new techniques to make this search tractable.
Existing file-system crash-consistency testing tools would not have found these bugs.
We also found that fuzzing was an effective way to invalidate assumptions
we had brought along from prior file systems experience,
such as the significance of unaligned writes and exercising per-CPU code paths.

Another assumption we carried into this work
was that the difficulty of building a PM file system
lies in correctly applying the PM programming model.
We intended to focus on exhaustively testing the precise persistency behavior of PM file system code.
However, we found instead that most PM file system bugs
were logic errors in optimizations that
could also have occurred in block-based file systems
(though likely would not improve performance in that context).
Existing tools that focus on detecting specific PM programming error patterns~\cite{Liu2019, Liu2020, pmemcheck, Liu2021, Di2021, Neal20, Fu2021, Gojiara2021}
would miss many of these bugs,
but we were able to detect them even with small workloads
and with very few in-flight writes applied.
Writing general-purpose consistency checks
and applying gray-box fuzzing to generate workloads
helped to invalidate these assumptions
and gave new insights into common bug patterns in PM file systems.


\end{lesson}

\begin{lesson}[Lightweight testing offers a scalable approach to detecting many crash-consistency bugs.]
\sysname is, in principle, a bounded exhaustive testing~\cite{Mohan18} (i.e., bounded verification) tool
for PM file systems:
given enough time, it can check every possible crash behavior
of every possible workload up to some bounds on its size and inputs.
Of course, it is not tractable to exhaust this search space even with very small bounds.
However, we found that \sysname is an effective \emph{lightweight} testing tool,
in that it can quickly and automatically find many bugs by checking small workloads and few crash states,
and then run for longer to find more corner-case issues.
\sysname runs the ACE seq-1 workloads on a file system in under 15 minutes
and, when considering only crash states with one persisted write,
can detect 12 of the \uniquebugs bugs we discovered.
On the other hand,
the fuzzer frontend to \sysname takes 1--2 orders of magnitude longer to run
(i.e., runs overnight)
but finds three more bugs than ACE.
These two frontends are complementary.
They enable a lightweight approach that helps developers iterate quickly on new code,
while still offering stronger confidence as the code gets ``closer to production''~\cite{bornholt:shardstore}.



\end{lesson}

\section{Conclusion}
\label{sec-conclusion}

This paper presents \sysname, a new record-and-replay framework for
testing the crash consistency of PM file systems. We use \sysname with
the ACE workload generator and the Syzkaller gray-box fuzzer and find
\totalbugs unique bugs across four PM file systems. To the best of our
knowledge, this is the largest corpus of crash-consistency bugs on PM
file systems. Our study of these bugs provides insights into how
crash-consistency bugs arise in PM file systems and what types of
tools are needed to test these systems.


\newpage

\bibliographystyle{plain}
\bibliography{references.bib}

\end{document}